\documentstyle[12pt]{article}
\newcommand{\s}[1]{\section{#1}\renewcommand{\theequation}
        {\mbox{\arabic{section}.\arabic{equation}}}\setcounter{equation}{0}}

\newcommand{\app}[1]{\section{#1}\renewcommand{\theequation}
        {\mbox{\Alph{section}.\arabic{equation}}}\setcounter{equation}{0}}
\renewcommand{\date}[1]{\par\bigskip\par\sl\hfill #1\par\medskip\par\rm}
\newcommand{\pacs}[1]{\smallskip\noindent{\sl PACS number(s):
                       \hspace{0.3cm}#1}\par\bigskip\rm}
\def\babs{\hrule\par\begin{description}\item{Abstract: }\it} 
\def\eabs{\par\end{description}\hrule\par\medskip\rm}
 
\renewcommand{\thanks}[1]{\footnote{#1}}

\renewcommand{\title}[1]{\begin{center}\Large\bf #1 \end{center}\rm\par\bigskip}
\renewcommand{\author}[1]{\begin{center}\Large #1\end{center}}
\newcommand{\address}[1]{\begin{center}\large #1 \end{center}}

\begin{document}

\newcommand{\reals}{\mbox{${\rm I\!R }$}}
\newcommand{\komplex}{\mbox{${\rm I\!\!\!C }$}}
\newcommand{\nats}{\mbox{${\rm I\!N }$}}
\newcommand{\intgs}{\mbox{${\rm Z\!\!Z }$}}

\newcommand{\can}{{\cal N}}
\newcommand{\cam}{{\cal M}}
\newcommand{\caz}{{\cal Z}}
\newcommand{\cao}{{\cal O}}
\newcommand{\cac}{{\cal C}}
\newcommand{\cah}{{\cal H}}

\newcommand{\al}{\alpha}
\newcommand{\be}{\beta}
\newcommand{\de}{\delta}
\newcommand{\ep}{\epsilon}
\newcommand{\ga}{\beta}
\newcommand{\la}{\lambda}
\newcommand{\th}{\theta}
\newcommand{\ze}{\zeta}

\newcommand{\De}{\Delta}
\newcommand{\Om}{\Omega}
\newcommand{\Si}{\Sigma}
\newcommand{\Ga}{\Gamma}

\newcommand{\rS}{{\rm S}}

\newcommand{\ent}{d (\nu )}
\newcommand{\pnenj}{\prod_{j=1}^p\nu_j}
\newcommand{\nn}{\nonumber}
\renewcommand{\theequation}{\mbox{\arabic{section}.\arabic{equation}}}
\newcommand{\intsi}{\int\limits_{\Sigma}d\sigma_x\,\,}
\newcommand{\back}{\bar{\Phi}}
\newcommand{\coba}{\bar{\Phi}^{\dagger}}
\newcommand{\abl}{\partial}
\newcommand{\PP}{{\rm PP\,\,}}
\newcommand{\pa}{\partial}
\newcommand{\qpi}{(4\pi)^{\frac{q+1} 2}}
\newcommand{\sul}{\sum_{l=-\infty}^{\infty}}
\newcommand{\snenp}{\sum_{n_1,...,n_p=0}^{\infty}}
\newcommand{\tint}{\int\limits_0^{\infty}dt\,\,}
\def\beq{\begin{eqnarray}}
\def\eeq{\end{eqnarray}}
\newcommand{\zb}{\zeta_{{\cal N}}}
\newcommand{\zbe}{\zeta_{{\cal N}+1}}
\newcommand{\rzb}{{\rm Res}\,\,\zb}
\newcommand{\Res}{{\rm Res\,\,}}
\newcommand{\fr}{\frac}
\newcommand{\sip}{\frac{\sin (\pi s)}{\pi}}
\newcommand{\rzs}{R^{2s}}
\newcommand{\g}{\Gamma\left(}
\newcommand{\ikma}{\int_{\gamma}\frac{dk}{2\pi i}k^{-2s}\frac{\pa}{\pa k}}
\newcommand{\suani}{\sum_{a=0}^i}
\newcommand{\zem}{\zeta_{{\cal M}}}
\newcommand{\hem}{A^{\cam}}
\newcommand{\hen}{A^{\can}}
\newcommand{\man}{{\cal M}}
\newcommand{\zeb}{\zeta_{{\cal B}}}
\newcommand{\pold}{D^{(d-1)}}
\newcommand{\zesd}{\zeta_{S^d}}
\newcommand{\fac}{\frac{(4\pi)^{D/2}}{a^d|S^d|}}
\newcommand{\facb}{\frac{(4\pi)^{d/2}}{a^d|S^d|}}
\newcommand{\zf}{{$\zeta$-function}}
\newcommand{\zfs}{{\zeta-functions}}
\newcommand{\hk}{K_{\can}^{1/2}(t)}
\newcommand{\pzb}{{\rm P}{\rm P}\,\,\zb}


\title{Heat-kernels and functional determinants on the      
generalized cone}
\author{
Michael Bordag\thanks{E-mail address: 
bordag@qft.physik.uni-leipzig.d400.de} and 
Klaus Kirsten\thanks{E-mail address:
kirsten@tph100.physik.uni-leipzig.de}}
\address{Universit{\"a}t Leipzig, Institut f{\"u}r Theoretische Physik,\\
Augustusplatz 10, 04109 Leipzig, Germany}
\author{Stuart Dowker\thanks{E-mail address:
dowker@a13.ph.man.ac.uk}}
\address{Department of Theoretical Physics,\\
The University of Manchester, Manchester M13 9PL, England}

\date{\today}
\babs
We consider zeta functions and heat-kernel expansions on the bounded, 
generalized cone in arbitrary dimensions using an improved calculational 
technique. The 
specific case of a global monopole is analysed in
detail and some restrictions thereby placed on the $A_{5/2}$ coefficient.
The computation of functional determinants is also addressed. General formulas 
are given and known results are incidentally, and rapidly, reproduced.
\eabs
\s{Introduction.}
In this paper we refine and generalise techniques developed 
earlier for the evaluation of heat-kernel expansion coefficients and functional
determinants of elliptic operators on manifolds with boundary. We concentrate 
on ball-like manifolds because precise answers can be found and,
apart from illustrating our method, the results for such specific 
manifolds are often useful in restricting the general forms of heat-kernel
coefficients. 

One of the motivations for this paper is to compute for a 
particular curved manifold whose boundary is not geodesically embedded. The
resulting restrictions are a little more informative than some others 
available \cite{kennedy78,kennedy78a,levitin}. The manifold also possesses a
singularity, which increases its interest.

For calculational simplicity the operator is taken to be the modified 
Laplacian, $\De-\xi R$, acting on scalars. The analysis could be extended
to forms without difficulty and also to other fields with a certain amount
of extra work 
\cite{esposito94,bar92,espo94,espo95,espo96,vassi95,vassipre}.   
It is possible that our techniques
will be of value in areas of physics where finite size systems
and boundary effects play a role, such as quantum cosmology and
statistical mechanics. 

In the next section we outline the geometry we have in mind and discuss 
the eigenmodes. The \zf\ is next constructed in section 3 and its 
properties translated into heat-kernel language in the following 
section. In order to make this paper reasonably self-contained the 
techniques alluded to previously are restated in improved and 
compactified form. The general method is applied to a global monopole in
section 5 and the results used in section 7 to place restrictions on the
numerical coefficients in a heat-kernel coefficient of some current 
mathematical interest \cite{bgv96}. Sections 9, 10 and 11 describe the 
evaluation of the functional determinant.

\s{The geometry and eigenmodes.}
The manifold in question can be termed the bounded, generalized cone and 
is defined as the $(d+1)$-dimensional space $\cam =I\times\can$ 
with the hyperspherical metric {\it cf} \cite{cheeger83}
\beq
ds^2=dr^2+r^2d\Si^2,  \label{2.1}
\eeq  
where $d\Si^2$ is the metric on the manifold $\can$, and $r$ runs from $0$ 
to $1$. $\can$ will be referred to as the base, or end, of 
the cone. If it has no boundary then it is the boundary of $\man$.

We note that the space is conformal to the product half-cylinder 
$\reals^+\times\can$,
\beq
ds^2=e^{-2x}\big(dx^2+d\Si^2),\quad x=-\ln r,\label{productform}
\eeq
which allows the curvatures on $\man$ and $\can$ to be related. The only
nonzero components of the curvature on $\man$ are, with obvious notation,
\beq
{R^{ij}}_{kl}={1\over r^2}\big({{{\widehat R}^{ij}}}_{\,\,\,\,kl}
-(\de^i_k\de^j_l-\de^i_l\de^j_k)\big),\quad\!\!
R^i_j={1\over r^2}\big(\widehat R^i_j-(d-1)\de^i_j\big),\quad\!\!
R={1\over r^2}\big(\widehat R-d(d-1)\big).
\label{curvature}
\eeq
These measure the local deviation of $\can$ from a unit $d$-sphere and 
indicate the existence of a singularity at the origin.
Finally, the extrinsic curvature is $\kappa^i_j=\de^i_j$ and we recognise 
(\ref{curvature}) at $r=1$ as the Gauss-Codacci equation.

Turning to analysis, the Laplacian is
\beq
\De_{\cam}={\pa^2\over\pa r^2}+{d\over r}{\pa\over\pa r}
+{1\over r^2}\De_{\can}.   \label{2.2}
\eeq
Boundary conditions are imposed at $r=1$ as will be described in the next 
section.

The nonzero eigenmodes of $\De_\cam$ that are finite at the origin have 
eigenvalues $-\al^2$ and are of the form
\beq
{J_{\nu}(\al r)\over r^{(d-1)/2}}\,Y(\Om), \label{2.3}
\eeq
where the harmonics on $\can$ satisfy
\beq
\De_{\can} Y(\Om)=-\la^2 Y(\Om)\label{2.4}
\eeq
and 
\beq
\nu^2= \la^2+(d-1)^2/4  . \label{2.5}
\eeq
It is easiest to allow for the addition of the term $-\xi R$ to 
$\De_\man$ when $\widehat R$ is constant and we shall assume that this is so 
in the detailed calculations presented later in this paper. The obvious 
example is the sphere, discussed {\it passim} in section 5. If we are 
interested 
solely in the Laplacian ($\xi=0$) this restriction is unnecessary.

The modes will still be as in equation (\ref{2.3}) with now
\beq
\nu^2=\la^2+(d-1)^2/4+\xi\big(\widehat R-d(d-1)\big)=\la^2
+\xi\widehat R+d(d-1)\big(\xi_d-\xi\big)
\label{index}
\eeq
where $\xi_d=(d-1)/4d$. For conformal coupling in $d+1$ dimensions, 
the last term disappears, as it also does when $d=0$ or $d=1$.

More generally, if $\widehat R$ is not constant, we write
$$
\De_{\man}-\xi R={\pa^2\over\pa r^2}+{d\over r}{\pa\over\pa r}
+{\xi d(d-1)\over r^2}+{1\over r^2}\big(\De_{\can}-\xi\widehat R\big)
$$
and introduce eigenfunctions, $\overline Y$, of the modified Laplacian 
on $\can$,
$$
\big(\De_\can-\xi\widehat R\big)\overline Y=-{\bar\la}^2\overline Y,
$$ 
so that the eigenfunctions on $\man$ are again of the form (\ref{2.3}) with
$Y$ replaced by $\overline Y$ and
\beq
\nu^2={\bar\la}^2+d(d-1)\big(\xi_d-\xi\big).\label{eigval}
\eeq
We assume that $\nu\ge1/2$ in order to avoid the appearance of types
of solution other than (\ref{2.3}).

\s{The zeta function on $\cam$.}
Let us first see how far the analysis can be taken without specifying the base
manifold $\can$. A boundary value problem may still easily be posed due to 
the form of the chosen metric. Both Dirichlet and generalized Neumann 
(or Robin) boundary conditions are to be considered and in the notation 
of, for example, \cite{levitin}, these read explicitly
\beq
J_{\nu}(\al) =0 \label{3.1a}
\eeq
for Dirichlet and 
\beq
\left(1-\frac D 2 -\ga \right) J_{\nu} (\al ) +\al J' _{\nu} (\al ) =0  
\label{3.1b} 
\eeq
for Robin. We set $D=d+1$ and use $D$ or $d$, whichever is convenient.

A handy way of organising eigenvalues is the 
Minakshisundaram-Pleijel \zf.
Let $d(\nu)$ be the number of linearly independent scalar harmonics on 
$\can$. Then the base zeta function is defined  by 
\beq
\zb (s) = \sum d(\nu) \nu^{-2s}=\sum d(\nu)
\big({\bar\la}^2+d(d-1)(\xi_d-\xi)\big)^{-s}\label{3.1}
\eeq
and our first aim will be to express the whole zeta function on $\man$, 
\beq
\ze_\man(s) = \sum \alpha ^{-2s},\nn
\eeq
as far as possible in terms of this quantity. That is, we seek to replace
analysis on the cone by that on its base in the manner of Cheeger 
for the infinite cone, \cite{cheeger83}.

We start with Dirichlet boundary conditions, the discussion for 
Robin conditions being virtually identical.

Following the analysis of \cite{bek,bgek} the starting point is the 
representation of the zeta function in terms of a contour integral
\beq
\zeta_{\cam} (s)= \sum \ent \ikma\ln
J_{\nu} (k),\label{3.2}
\eeq
where the anticlockwise contour $\gamma$ must enclose all the solutions
of (\ref{3.1a}) on the positive real axis (for a similar treatment of
the zeta function as a contour integral see
\cite{kamenshchikmishakov92,barvinskykamenshchikkarmazin92,bordag95}).

As was seen in \cite{bek,bgek} it is very useful to split the 
zeta function into two parts. To actually perform this separation, 
some notation for the uniform asymptotic expansion of the 
Bessel function 
$I_{\nu} (k)$ is required. For $\nu \to \infty$ with $z=k/\nu$
fixed one has, \cite{olver54,abramowitzstegun72},
\beq
I_{\nu} (\nu z) \sim \frac 1 {\sqrt{2\pi \nu}}\frac{e^{\nu
\eta}}{(1+z^2)^{\frac 1 4}}\left[1+\sum_{k=1}^{\infty} \frac{u_k (t)}
{\nu ^k}\right],\label{3.3}
\eeq
with $t=1/\sqrt{1+z^2}$ and $\eta =\sqrt{1+z^2}+\ln
\big(z/(1+\sqrt{1+z^2})\big)$.
The first few coefficients are listed in \cite{abramowitzstegun72}.
Higher coefficients are immediately obtained by using the recursion
\cite{olver54,abramowitzstegun72} 
\beq
u_{k+1} (t) =\frac 1 2 t^2 (1-t^2) u'_k (t) +\frac 1 8 \int\limits_0^t
d\tau\,\, (1-5\tau^2 ) u_k (\tau ),\nn
\eeq 
starting with $u_0 (t) =1$. As is clear, all the $u_k (t)$ are
polynomials in $t$. We also need the 
coefficients $D_n (t)$ defined by the cumulant expansion
\beq
\ln \left[1+\sum_{k=1}^{\infty} \frac{u_k (t)}{\nu ^k}\right] \sim
\sum_{n=1}^{\infty} \frac{D_n (t)}{\nu ^n}\label{3.4}
\eeq
which have the polynomial structure
\beq
D_n (t) = \sum_{b=0}^n x_{n,b}\, t^{n+2b}.\label{3.5}
\eeq
From the small $z$ behaviour of eq.(\ref{3.3}) one derives the important
value $D_n(1)=\ze_R(-n)/n$, which will be seen later on in 
eq.~(\ref{9.2}).

By adding and subtracting $N$ leading terms of the asymptotic
expansion, eq.~(\ref{3.4}), and performing the same steps as described
in \cite{bek,bgek} one finds the split
\beq
\zeta_{\cam} (s) =Z(s) +\sum_{i=-1}^{N}A_i(s),\label{3.6}
\eeq
with the definitions
\beq
Z(s) &=& \sip \sum \int\limits_0^{\infty}
dz\,\,(z\nu )^{-2s}\frac{\pa}{\pa z}
\bigg(\ln\left(z^{-\nu}I_{\nu} (z\nu
)\right)\label{3.7}\\
& &\hspace{3cm}
-\ln\bigg[\frac{z^{-\nu}}{\sqrt{2\pi\nu}}\frac{e^{\nu\eta}}
{(1+z^2)^{\frac
1 4}}\bigg]-\sum_{n=1}^N \frac{D_n (t)}{\nu ^n}\bigg),\nn
\eeq
and 
\beq
A_{-1} (s) &=& \frac 1 {4\sqrt{\pi}} \frac{\Gamma \left(s-\frac 1 2\right)}
{\Gamma (s+1)} \zb \left( s-1/2 \right),\label{3.8}\\
A_0 (s) &=& -\frac 1 4 \zb (s), \label{3.9}\\
A_i (s) &=& -\frac 1 {\Gamma (s)} \zb \left( s+i/2 \right)
\sum_{b=0}^i x_{i,b} \frac{\g s+b+i/2\right)}
{\g b+i/2\right)}.\label{3.10}
\eeq
The function $Z(s)$ is analytic on the strip $(1-N)/2 < \Re s$, which may
be seen by considering the asymptotics of the integrand in eq.~(\ref{3.7})
and by having in mind that the $\nu^2$ are eigenvalues of a second order
differential operator, see eq.~(\ref{2.5}).


As is clearly apparent in eq.~(\ref{3.8})--(\ref{3.10}), base contributions 
are separated from radial ones.
We will see in the following section that this enables 
the heat-kernel coefficients of the Laplacian on the manifold $\cam$ to be
written in terms of those on $\can$.

In order to treat Robin boundary conditions, only a few changes are necessary.
In addition to expansion (\ref{3.3}) we need
\cite{olver54,abramowitzstegun72}
\beq
I_{\nu} ' (\nu z )\sim 
\frac 1 {\sqrt{2\pi \nu}} \frac{e^{\nu \eta} (1+z^2)^{1/4}} z 
\left[1+\sum_{k=1}^{\infty}\frac{v_k(t)}{\nu^k}\right], \label{3.11}
\eeq
with the $v_k (t)$ determined by
\beq
v_k (t) = u_k (t) +t (t^2 -1) \left[
\frac 1 2 u_{k-1} (t) +t u_{k-1} ' (t) \right].\label{3.12}
\eeq
The relevant polynomials analogous to the $D_n (t)$, eq.~(\ref{3.4}), are 
defined by
\beq
\ln\left[ 
1+\sum_{k=1}^{\infty}\frac{v_k (t)}{\nu^k} +
\frac{ 1-D/2 -\ga }{\nu} t \left(1+\sum_{k=1}^{\infty} \frac{u_k (t)}
{\nu^k}\right) \right]
\sim  \sum_{n=1}^{\infty} \frac{M_n (t)}{\nu^n}\label{3.13}
\eeq
and have the same structure,
\beq
M_n (t) =\sum_{b=0}^n z_{n,b}\, t^{n+2b}.\label{3.14}
\eeq

One may again introduce a split as in eq.~(\ref{3.6})
with $A_{-1}^R (s) =A_{-1} (s)$, $A_0^R(s) =-A_0 (s)$ and eq.~(\ref{3.10})
remains valid when $x_{i,a}$ is replaced by $z_{i,a}$.

\s{Heat-kernel coefficients on the generalized cone.}
The previous formulas are already sufficient to give the heat-kernel
coefficients on the manifold $\cam$ in terms of those on 
$\can$. However, before giving the relation, some special circumstances 
of our situation must be explained and, 
for expository purposes, a conventional short-time expansion of a generic
heat-kernel will now be displayed,
\beq
K (t)\sim \sum_{n=0,1/2,1,...}^{\infty} C_n\, t^{n-D/2}.
\label{ex}
\eeq

For a geometric operator, such as the Laplacian on a Riemannian 
manifold, the coefficients $C_n$ are integrals of polynomials in the
curvatures and their derivatives. Typically, the volume integrand of $C_n$ 
contains terms $\sim R^n$ and it can be seen from the form of the 
curvature in the present case, (\ref{curvature}), that the naively integrated
coefficients $C_n$ for $n\ge D$ diverge. For this range, Cheeger 
\cite{cheeger83} and Br\"uning \cite{bruning84} show that the relevant 
quantity in (\ref{ex}) is the {\it partie finie} of the 
integral. This is obtained, in our case, by restricting the radial 
integral to $\ep\le r\le1$, $\ep>0$, and taking the finite remainder as 
$\ep\to0$. An equivalent procedure for a given $n$ is to evaluate in a 
dimension $D>n$ and then continue to the required dimension, {\it \`a la}
dimensional regularisation.

A further aspect of singular problems is the existence of logarithmic
terms in the heat-kernel expansion, \cite{bruningseeley87}. We introduce 
these via the important \zf\ value $\zeta(0)$ which is finite for a 
nonsingular elliptic operator on a smooth manifold. In the present case,
(\ref{3.6}) and (\ref{3.8}) show that $\ze_\man(s)$ has a pole at $s=0$
provided $\ze_\can(s)$ has one at $s=-1/2$. According to a standard relation,
the residue is proportional to the heat-kernel coefficient  
$A^\can_{(d+1)/2}$ on $\can$, and so, if $\can$ is closed, vanishes for 
even $d$, being then a pure boundary term.
A pole at $s=0$ in the \zf\ translates into a `nonstandard' logarithm
term in the heat-kernel expansion as we now show.

Let $K_\cam(t)$ be the heat-kernel associated with the modified Laplacian on 
$\cam$, and define now the coefficients by the convention
\beq
K_{\cam } (t) \sim \sum_{n=0,1/2,1,...}^{\infty} A_n^{\cam}\,t^{n-D/2}
+A'\,\log t.
\label{4.1}
\eeq
A Mellin transform argument relating the 
heat-kernel to the \zf, going back to \cite{MandP49}, see also {\it eg}
\cite{voros87}, gives 
\beq
\hem _{n/2} = \Res\zem (s) \Gamma (s)\bigg|_{s=(D-n)/2},
\quad n\ne D,\label{4.2}
\eeq
while
\beq
\hem_{D/2}=\PP\zem(0),\quad {\rm and}\,\,A'=- \Res\zem(0).
\eeq
From (\ref{3.8})--(\ref{3.10}) we find in particular
\beq
\PP\zem(0)&=&(\ln2-1)\,\Res\ze_\can(-1/2)-{1\over2}\PP\ze_\can(-1/2)
-{1\over4}\ze_\can(0) \nn\\
& &-\sum_{i=1}^{D-1}{1\over i}\ze_R(-i)\,\Res\ze_\can(i/2)
\label{zeta0}
\eeq
and
\beq
\Res\zem(0)=-{1\over2}\Res\ze_\can(-1/2).
\eeq
The singularity evidences itself in contributions to the constant and
logarithmic terms in the expansion, {\it cf} \cite{cheeger83}.

In the later calculations it will be arranged that the logarithmic term 
does not occur. Its absence permits a standard evaluation of the
functional determinant. 

We consider an arbitrary dimension, $D$, and so, in practice,
it will be enough to work with $n<D$ in order to determine {\it any} 
coefficient. In consequence we use
\beq
\hem _{n/2}=
\Gamma\big( (D-n)/ 2 \big)\,\Res\zem \big( (D-n)/ 2 \big)
,\label{4.3}
\eeq
in the following and continue in $D$ as described above. 

Denoting by $\hen_ n$ the heat-kernel
coefficients associated with $\zb$ by the corresponding equations, we may 
write as an immediate consequence of eqs.~(\ref{3.8})--(\ref{3.10}) and
(\ref{4.3}) the basic relation,
\beq
\hem _{n/2} &=&
 \frac 1 {2\sqrt{\pi} (D-n)} \hen _{n/2} -\frac 1 4 \hen _{(n-1)/2} \label{4.4}
\\
& &-\sum_{i=1}^{n-1} \hen _{(n-1-i)/2} \sum_{b=0}^i x_{i,b}\, \frac
{\Gamma\left((D-n+i)/2 +b\right)}
{\Gamma \left((D-n+i)/2\right)\Gamma(b+ i/2)}\nn
\eeq
with $\hen _{(n-1)/2}=0$ for $n=0$. Thus, given the coefficients on $\can$, 
eq.~(\ref{4.4}) relates them immediately to the coefficients on $\cam$.
The boundary condition at $r=1$ is encoded just in the sum over $b$.
This relation can be used to restrict the general form of the 
heat-kernel coefficients as will be explained briefly in sections 6 and 7.

Eq.~(\ref{4.4}) remains true for Robin conditions once the sign of the 
second term on the right-hand side is reversed and the $x$'s replaced 
with the $z$'s.
\s{The global monopole.}
Let us illustrate this formalism with a bounded version of the simplified 
global monopole introduced by Sokolov and Starobinsky 
\cite{sokolovstarobinsky77} and discussed more physically by Barriola and 
Vilenkin \cite{barrvil89}. 

The manifold $\can$
is a $d$-sphere of radius $a$ and is the boundary $\pa\man$ of 
$\man$ at $r=1$. If $a$ is not unity, this produces
a distortion which exhibits itself as a solid angle deficit, or excess,
at the origin. The $d+1$-space is not flat, unless $a=1$, having scalar
curvature 
\beq
R=d(d-1){1-a^2\over a^2r^2}.\label{4.5}
\eeq

There are two useful conformal transformations. The first takes the
metric into a product form (as in (\ref{productform}))
$$
ds^2=dr^2+a^2r^2d\Om^2=e^{-2x}\big(dx^2+a^2d\Om^2),\quad x=-\ln r
$$
which is that of the (Euclidean) Einstein universe, 
$\reals^+\times\rS^d_a$, where
$\rS^d_a$ is a sphere of radius $a$. This is conformally flat and so, 
therefore, is the monopole metric, a direct statement being
$$
dr^2+a^2r^2d\Om^2=\big({r\over l}\big)^{2\al}\big(dr^2+r^2d\Om^2)
$$
where $\al$ is given by ($d\ne0,1$)
$$
\al=1-\big(1+(1-a^2)/a^2\big)^{1/2}
$$
and $l$ is an arbitrary scale parameter. This is reminiscent of a 
Schwarz-Christoffel transformation, the conformal nature of the map
breaking down at the origin.

The first conformal relation gives the nonzero curvature components in
terms of those on the unit sphere, (see (\ref{curvature})),
\beq
{R^{ij}}_{kl}={1-a^2\over a^2r^2}\,
\big(\de^i_k\de^j_l-\de^i_l\de^j_k\big).\label{4.6}
\eeq

As explained earlier, the mode decomposition goes through exactly as 
in the flat case except that the order of the Bessel function acquires 
an extra shift,
$$
\nu^2={\la^2\over a^2}+{(d-1)^2\over4}+\xi d(d-1){1-a^2\over a^2} 
$$ where $\la^2$ are the eigenvalues of the Laplacian on the unit 
$d$-sphere. Hence (see (\ref{index}))
$$
\nu^2={(n+(d-1)/2)^2\over a^2}+d(d-1){1-a^2\over a^2}
\big(\xi-\xi_d\big)
$$
so that if $\xi=\xi_d$ we obtain the usual 
simplification ({\it cf} \cite{mazzitellilousto91} for $d=2$)
$$
\nu={1\over a}\big(n+{d-1\over2}\big),\quad n=0,1,2\ldots,
$$
and the base $\zeta$-function is given by a 
simple scaling of the unit sphere $\zeta$-function,
which is the one appropriate for the uncompressed ball,
\beq
\ze_{\can}(s)= a^{2s}\ze_{\rS^d}(s).\label{4.7}
\eeq

A point to note is that there is no pole at $s=0$ in 
$\ze_{\man}(s)$. As stated, this can only possibly occur if $d$ is odd 
(for a closed $\can$) and we know that the $\ze$-function (\ref{4.7}) is a 
finite sum of Riemann $\ze$-functions and has no pole at $s=-1/2$. This is 
actually a consequence of our choice of conformal coupling.

In order to apply eq.~(\ref{4.4}) to the monopole, the residues
of the base zeta function, $\zb$, are needed. These may be obtained most 
easily using its representation in terms of the Barnes zeta function
\cite{barnes03} defined by the sum
\beq
\zeb (s,a|\vec r ) &=& \sum_{\vec m =0}^{\infty} \frac 1 {(a+\vec m. \vec r )
^s},\label{4.9}
 \eeq
valid for $\Re s > d$, with the $d$-vectors $\vec m$ and $\vec r$. 

It is shown in \cite{changdowker93} that the zeta 
function on a portion of the $d$-sphere determined by the degrees
$\vec r$ corresponds to the value $a=\sum r_i-(d-1)/2$ for Dirichlet 
and to $a=(d-1)/2$ for Neumann conditions on its perimeter. However we do not 
need the full generality of these statements and can limit ourselves to the 
case $\vec r=\vec 1$ corresponding to the hemisphere,
\beq
\zeb (s,a|\vec 1 ) &=& \sum_{l=0}^{\infty} \left(
\begin{array}{c}
l+d-1\\ d-1 \end{array}
\right)
\frac 1 {(a+l)^s}.
 \eeq

It is easily shown, algebraically, that the full-sphere zeta function 
is the sum of the hemisphere Dirichlet and Neumann functions, 
\cite{dowker94}. 
To see this, remember that the number of independent scalar harmonics on 
$\can=\rS^d$ is 
\beq
d(l) = (2l+d-1) \frac{(l+d-2)!}{l!\,(d-1)!}.\label{4.8}
\eeq
For reasons of space, we do not give any history of 
sphere zeta functions.

Using the notation 
$\zeb (s,a|\vec 1 ) =\zeb (s,a)$ one finds
\beq
\zesd (s) =\zeb\big( 2s, (d+1)/2\big) 
+\zeb \big(2s,(d-1)/2\big), \label{4.10}
\eeq
which reduces the analysis of the sphere zeta function to that of the 
Barnes function. Using the integral representation
\beq
\zeb (s,a) =\frac {i\Gamma (1-s)}{2\pi} 
\int_L dz\,\,\frac{e^{z\left( d/ 2 -a\right)} (-z)^{s-1}}
{2^d\sinh^d\big(z/2\big)},\label{4.11}
\eeq
where $L$ is the Hankel contour, one immediately finds for the base 
function
\beq
\zb (s) &=& a^{2s}  \frac{i\Gamma (1-2s)}{2\pi} 2^{2s+1-d} \int_L
dz  \,\, (-z)^{2s-1}\frac{\cosh z}{\sinh ^d z}\label{4.12}\\
&  =&a^{2s}\frac{i\Gamma(2-2s)}{2\pi(d-1)}2^{2s+1-d}\int_L
dz \,\,{(-z)}^{2s-2}{1\over\sinh^{d-1}z}.\nn
\eeq
For the residues this yields ($m=1,2,...,d$)
\beq
\Res\zb (m/2) = a^{m}\frac{2^{m-d} \pold _{d-m}}{(d-1)(m-2)! (d-m)!} ,
\label{4.13}
\eeq
with the $\pold _{\nu}$ defined through ({\it cf} \cite{chodosmyers84})
\beq
\left(\frac z {\sinh z}\right)^{d-1} =\sum_{\nu =0}^{\infty} \pold _{\nu}
\frac{z^{\nu}}{\nu !}.\label{4.14}
\eeq
Obviously $\pold _{\nu}=0$ for $\nu$ odd, so there are actually poles only for
$m=1,2,...,d$ with $d-m$ even. 
The advantage of this approach is that known recursion formulas allow 
efficient evaluation of the $D^{(n)}_\nu$ as polynomials in $d$, 
\cite{norlund22}. 

Using eq.~(\ref{4.13}) in eq.~(\ref{4.4})
we find for the heat-kernel coefficients $\hem_ {k/2}$ 
\beq
\frac{(4\pi)^{D/2}}{|S^d|}
\hem _{k/2} &=& \frac{(d-k-1)}{(d-1)(d-k+1) k!}\left( 
\frac{d+1-k} 2\right)_{\displaystyle{k/2}} \pold _k a^{d-k}\nn\\
& &-\frac{(d-k)}{4(d-1)(k-1)!} \left(
\frac{d+2-k} 2 \right) _{\displaystyle{(k-1)/2}}\pold _{k-1} a^{d+1-k}\nn\\
& &-\frac{2\sqrt{\pi}} {(d-1)} \sum_{i=1}^{k-1}
\frac{d+i-k}{(k-1-i)!}\left(\frac{d+2-k+i} 2 \right)_{\displaystyle{
(k-i-1)/2}}\times\label{4.15}\\
& &\qquad \sum_{b=0}^i\frac{x_{i,b}}{\Gamma\left( b +i/2\right)}
\left(\frac{d+1-k+i} 2 \right)_{\displaystyle{b}} a^{d+1+i-k},\nn
\eeq
where $(y)_n =\Gamma (y+n)/\Gamma (y)$ is the Pochhammer symbol.
Eq.(\ref{4.15}) exhibits the heat-kernel coefficients as explicit functions 
of the dimension $d$ and, although derived for $k<D$, they can now be extended
beyond this range.

For $a=1$ (\ref{4.15}) reduces to the coefficients on the ball and is in
full agreement with the results of Levitin \cite{levitin}. The 
polynomials up to $A_3^{\cam}$ are listed in Appendix A.

For Robin boundary conditions one has to make the modifications outlined
at the end of section 4. The results are summarized in Appendix B, once more
up to $\hem _3$.

\s{Comparison with usual expressions.}

The intention is to put restrictions on general forms of the $A^\man_{k/2}$ 
using the particular results for the monopole. There is however the possible
problem of a contribution from the singularity at the origin.
Does a piece have to be added specially to
the usual forms to account for this? The effect of the singularity 
appears only in the constant and logarithmic terms in the heat-kernel 
expansion and so
only $A^\man_{D/2}$ is affected. However the calculation provides unique
polynomials in $D$ for all $k$. Does anything special happen for $k=D$?
We show that it does and that singularity terms do not have to
be added to the usual smooth forms. An example will illustrate the 
general point. 

For $D=2$ the compressed
ball is an ordinary cone of angle $2\pi a$. Consider now the usual 
Dirichlet smeared expression for a {\it smooth} $D$-manifold
\beq
(4\pi)^{D/2}A_1(f)={1\over6}(1-6\xi)\int_{\man}Rf
+{1\over3}\int_{\pa\man}\big(\kappa f-{3\over2}n.\pa f\big)
\label{ayonesmooth}
\eeq
and, to avoid the log term, set $\xi=\xi_d=(d-1)/4d$. Substituting $R$ from 
(\ref{4.5}) and $\kappa=d$, (\ref{ayonesmooth}) becomes, on the 
compressed ball,
\beq
A_1(f)=a^d\bigg((3-d){1-a^2\over12a^2}\sum_{j=0}^\infty
{(d-1)f^{(2j)}(0)\over(d-1+2j)2j!}
+{d\over3}f(1)+{1\over2}f'(1)\bigg){|\rS^d|\over(4\pi)^{D/2}},
\label{ayoneball}
\eeq
where we have assumed that the smearing function $f$ depends on $r^2$ only.

Note that the $(d-1)$ factor, making $R$ vanish on the disc,
cancels against a corresponding factor from the integration over
$r$ for $j=0$ so that the volume term remains nonzero at $D=2$. Then, 
evaluating at $D=2$ gives
\beq
A_1(f)={1\over12}\bigg({1-a^2\over a}f(0)+2af(1)\bigg)+{a\over2}f'(1)
\label{ayonetwo}
\eeq
which can be compared with the standard expression for a $2$-manifold 
with a conical singularity of angle $\be$ at the origin,
\beq
A_1(f)={1\over24\pi}(1-6\xi)\int_{\man}Rf
+{1\over12\pi}\int_{\pa\man}(\kappa f-{3\over2}n.\pa f)
+{1\over12}\bigg({2\pi\over\be}-{\be\over2\pi}\bigg)f(0),
\label{ayonecon}
\eeq
usually derived from the Sommerfeld-Carslaw heat-kernel on the 
cone. (It can be generalised to any dimension.)

Evaluated directly on the compressed 2-ball where $R=0$ now and 
$\kappa=1$, (\ref{ayonecon})
agrees with (\ref{ayonetwo}). We see that the singularity part of 
(\ref{ayonecon}) arises as 
the $D\to2$ limit of the volume integration over the monopole 
curvature densities in the usual {\it smooth} expression. In this way the 
detailed analysis of the cone heat-kernel could be avoided. This is also
true if $\xi\ne\xi_d$ because of the $(d-1)$ factor in the eigenvalues,
eqs. (\ref{index}) or (\ref{eigval}).

The $D=4$ case can be investigated in a similar fashion by examining $A_2$.
%
%
%
%
In the general case, a value being fixed for $k$, the volume integrand of 
$A_{k/2}$ (if there is one) vanishes at $D=k$ because of the conformal
flatness. However the {\it integrated} volume
$A_{k/2}$ remains nonzero and is the contribution of the singularity. 

Another application of the $A_2$ coefficient presents itself. For the ordinary
cone, it is known that the smeared heat-kernel expansion consists
of a series of rational functions in the apex angle which
are straightforwardly calculated as residues from the Sommerfeld-Carslaw 
expression. The singularity term in (\ref{ayonecon}) is the first of 
these functions. The second will come from the smeared $A_2$
evaluated at $D=2$. If $f(r^2)$ is Taylor expanded about the origin, because 
of the factor $(d-1)$ in (\ref{4.5})
all terms in $f$ in the volume part will vanish except that proportional 
to $r^2$ which yields a factor of $(d-1)$ in the denominator giving
a nonzero result. This would allow one to obtain the second of these 
residue functions, although this is not the best way. The upshot is that the 
conical singularity in two dimensions can be exactly simulated by a monopole 
in $D$ dimensions as the $D\to2$ limit of the smooth formulation.

The main conclusion of this section is that the polynomial forms deduced in
the present paper for the monopole can be compared immediately with the 
usual smooth general forms, in so far as these are known. This we proceed 
to do in a particular case.
\s{The $A_{5/2}$ coefficient.}


Branson, Gilkey and Vassilevich \cite{bgv96} (Lemma 5.1) give the 
general form of this coefficient and determine many of the numerical
coefficients. Looking at the $A_{5/2}$ expression for the global
monopole, section 5 and Appendices A and B, we are able to fix some
additional numbers. We do not give here a comprehensive
treatment of this question, intending only that it should illustrate
our general method. It is possible that the restrictions could be found
by easier means.

Using Lemma 5.1 of ref.~\cite{bgv96} for the global monopole, inserting the
geometric tensors given in the previous sections and comparing with the
polynomials in Appendices A and B, we find for Dirichlet boundary conditions,
\beq
d_{36}^-=-\frac{65}{128};\qquad d_{37}^- =-\frac{141}{32};
\qquad d_{40} ^- =-\frac{327} 8, \nn
\eeq
together with the relations,
\beq
d_{38}^-+d_{39}^- &=&1049/32,\nn\\
d_1^-+2d_{27}^- -2d_{29}^- &=& -504,\nn\\
d_1^--4d_2^--2d_{25}^- &=& -360.
\eeq
For Robin boundary conditions the results read
\begin{equation}
\begin{array}{rclrclrcl}
d_{21}^+ &=& -60; & d_{30}^+ &=& 2160; & d_{31}^+&=& 1080;  \\
d_{32}^+ &=& 360; & d_{33}^+ &=& 885/4; & d_{34}^+ &=& {315}/2; \\
d_{35}^+ &=& 150; 
& d_{36}^+ &=& {2041}/{128}; & d_{37}^+ &=& {417}/{32};\\
d_{40}^+ &=& {231}/8, &    & &    &    & &    
\end{array}
\end{equation}
with the additional relations,
\beq
d_{38}^  ++ d_{39}^+ &=&1175/32,\nn\\
d_1^++2d_{27}^+ -2d_{29}^+ &=& 186 ,\nn\\
d_1^+-4d_2^+-2d_{25}^+ &=& -130.
\eeq
For Dirichlet conditions our example thus reduces the 
number of unknown numerical coefficients effectively by $6$, and for Robin 
by $13$.

Eq.~(\ref{4.4}) also allows one to place restrictions on the general form of
the coefficients which we want to describe briefly. Assume that $\can$
is closed and thus has no boundary so $\pa \cam =\can$ and $\hen _{n/2} 
=0$ for $n$ odd. The idea will already be clear from the lowest example $n=1$.
Then eq.~(\ref{4.4}) gives 
\beq
\hem _{1/2} = -\frac 1 4 \hen _0 = -\frac 1 4 (4\pi)^{(D-1)/2}\, 
|\pa \cam|,
\nn
\eeq 
which is of course the known result. As a rule, knowing the volume coefficient
$A^\can_n$, relation (\ref{4.4}) puts restrictions on the coefficient
$A^\man_{n+1/2}$. Let us illustrate this a bit more.
Choosing the operator $-(\Delta +E )$, $E$ depending only on the coordinates
on $\can$, $A^\man_{3/2}$ has the structure (we use no smearing function 
this time)
\beq
A^\man_{3/2} =-(384)^{-1} (4\pi )^{(D-1)/2} \int_{\pa \cam}
(d_1 E+d_2 R+d_3 R_{ab}\, n^an^b+d_4 \kappa ^2 +d_5 \kappa_{ab} \kappa^{ab})
.\nn
\eeq
Employing eq.~(\ref{4.4}) again,
\beq
\hem _{3/2} =-\frac 1 4 \hen _1 -{1\over64}\hen _0 (11-12D+5D^2),\nn
\eeq
and also the Gauss-Codacci relations between the intrinsic and ambient 
geometries of $\can$, one finds the known numbers,
\beq
d_1 =96;\qquad d_2 =16 ; \qquad d_4 =7; \qquad d_5 =-10,\nn
\eeq
just from a knowledge of the volume term $\hen _1$.
Several numerical coefficients might also
be determined for the $A^\man_{5/2} $ coefficient in this way but, taking the 
computations of ref.~\cite{bgv96}
and the previous results together, probably nothing new would emerge.

\s{Lens space bases.}
Examples of locally spherical bases, $\can$, are the homogeneous spaces
$\rS^d/\Gamma$ where $\Gamma$ is a discrete group of free isometries of 
$\rS^d$. The corresponding infinite cone ($0\le r<\infty$) has been 
treated in \cite{dowker89}. For simplicity the sphere radius is set to
unity to make the cones locally flat and we consider only the value
$\ze_\man(0)$ in detail. Because of the homogeneity, the heat-kernel 
coefficients on $\rS^d/\Gamma$ are simply a factor of $|\Gamma|$ smaller
than those on the unidentified sphere and, therefore, so are those on 
$\man$ computed according to (\ref{4.4}).

For even $D$, equation (\ref{zeta0}) reduces to
\beq
\ze_\man(0)=-{1\over2}\ze_\can(-1/2)
-\sum_{i=1}^{d}{1\over i}\ze_R(-i)\Res\ze_\can(i/2)\label{zeta0b}
\eeq
the first part of which we recognise as being the negative of the total 
Casimir energy on $\can$. 
The second part is written by homogeneity in terms of the full sphere
value
\beq
\sum_{i=1}^{d}{1\over i}\ze_R(-i)\Res\ze_{\rS^d/\Ga}(i/2)=
{1\over|\Ga|}\sum_{i=1}^{d}{1\over i}\ze_R(-i)\Res\ze_{\rS^d}(i/2).
\eeq

To obtain the contribution due solely to the singularity, $1/|\Gamma|$ of
the complete sphere value must be subtracted from (\ref{zeta0b}). We see
that the last term cancels and so
\beq
\ze^{{\rm sing}}_\man(0)=-{1\over2}\bigg(\ze_{\rS^d/\Ga}(-1/2)
-{1\over|\Ga|}\ze_{\rS^d}(-1/2)\bigg).
\eeq

As a simple case consider the three-dimensional lens space 
($\Ga=\intgs_m$). Then 
\beq
\ze^{{\rm sing}}_\man(0)={1\over720m}(m^2-1)(m+11)
\eeq
using the Casimir energy calculated in \cite{dowkerbanach78}. This agrees
of course with the value in \cite{dowker89}. The higher-dimensional cases,
and other groups, can be treated by various means.
\s{Dirichlet functional determinants.}
As a further application of the ideas presented in the previous sections, let
us consider the functional determinant on the generalized cone. To avoid 
problems of definition, we must assume that $A^\can_{(d+1)/2}$ vanishes. 
This eliminates the
possibility of a pole in $\ze_\man(s)$  at $s=0$ and the determinant is
then conventionally defined to be $\exp\big(-\ze_\man'(0)\big)$.

For the calculation of the determinant we have seen in \cite{bgek} that 
the first $D -1 =d$ terms in the asymptotic expansion are to be removed. 
Thus, from now on, we
set $N=d$ in eq.~(\ref{3.6}). The contribution of the $A_i$'s to the 
determinant may be immediately given,
\beq
A_{-1}' (0) &=& (\ln 2 -1)\,\zb (-1/2)  -\frac 1 2 \zb ' (-1/2),\nn\\
A_0 ' (0) &=& -\frac 1 4 \zb ' (0),\label{9.1}\\
A_i ' (0) &=& -\frac{\zeta_R (-i)} i \bigg( \gamma \rzb (i/2) 
 + \pzb (i/2)\bigg)\nn\\
& & -\sum_{b=0}^i x_{i,b}\, \psi (b+i/2)\,\rzb (i/2),\nn
\eeq
with $\psi (x) =(d/dx)\ln\Gamma (x)$.
Following the procedure in \cite{bgek} we find 
\beq
Z' (0) =\sum d(\nu ) \int_0^{\infty}dt\, 
\left(\sum_{n=1}^d \frac{D_n (1) }{(n-1)!}t^n +\frac 1 2 -\frac 1 t +
\frac 1 {e^t-1} \right) \frac{e^{-t\nu}} t, \label{9.2}
\eeq
which is well defined by construction, as is seen explicitly using the 
small $t$ asymptotic expansion,
\beq
\frac 1 {e^t-1} = \frac 1 t -\frac 1 2 -\sum_{n=1}^{\infty} \frac{t^n} {n!}
\,\zeta_R (-n),\label{9.3}
\eeq
and the value $D_n (1) = \zeta_R (-n) /n$. This expansion may also be 
used to obtain a kind of asymptotic series for $Z' (0)$,
\beq
Z' (0) =\sum_{n=d+1} ^{\infty} \frac{\zeta_R (-n)} n \zb (n).\nn
\eeq
However, as a rule, $\zb (n)$ can be determined only numerically once the 
eigenvalues are known.

Introducing the `square root' heat-kernel associated with the eigenvalue 
$\nu$, 
\beq
\hk =\sum d(\nu) e^{-t\nu} ,
\nn
\eeq
eq.~(\ref{9.2}) can be written in the form
\beq
Z' (0) = \int_0^{\infty}dt\, \frac 1 t 
\left(\sum_{n=1}^d \frac{D_n (1) }{(n-1)!}t^n +\frac 1 2 -\frac 1 t +
\frac 1 {e^t-1} \right)   \hk .\nn
\eeq
Let us calculate the individual pieces, as far as possible.
For this purpose, introduce a regularisation parameter, $z$, and define
\beq
Z' (0,z) = \int_0^{\infty}dt\, t^{z-1 }  
\left(\sum_{n=1}^d \frac{D_n (1) }{(n-1)!}t^n +\frac 1 2 -\frac 1 t +
\frac 1 {e^t-1}\right)   \hk .\nn
\eeq
This leads to 
\beq
Z' (0,z) &=& \sum_{n=1}^d \frac{D_n (1)}{(n-1)!} \Gamma (n+z)\, 
\zb \left( \frac{z+n} 2 \right) +\frac 1 2 \zb \left(\frac z 2 \right) 
\Gamma (z) \label{9.4}\\
& &-\zb \left(\frac{z-1} 2 \right) \Gamma (z-1) 
+\ze_{\can+1}(z)\Ga(z),\nn
\eeq
which we need at $z=0$.
Here we have introduced, as seems natural, the zeta function
\beq
\zeta_{\can+1}(z)
=\sum_{n=1}^{\infty} \sum d(\nu ) (\nu +n)^{-z}
={1\over\Gamma(z)}\sum d(\nu ) 
\int_0^{\infty} dt\,\, t^{z-1} \frac{e^{-t\nu}}{e^t-1}.\label{nzet}
\eeq

Eq.~(\ref{9.4}) may be expanded around
$z=0$ and the required derivative expressed in the intermediate form 
\beq
\zeta_{\cam}' (0) &=& \sum_{i=1}^d  \rzb (i/2)
\left( 
\frac{\zeta_R (-i)} i\left(
-\gamma +2\psi (i)\right) -\sum_{b=0}^i x_{i,b}\, \psi (b+i/2) \right)\nn\\
& &-\frac 1 2 \gamma \zb (0) +(\ln 2 -\gamma )\, \zb (-1/2) \label{9.5}\\
& &+\lim_{z\to 0}\left(\sum_{i=1}^d \frac 2 {zi} \zeta_R (-i)\, \rzb (i/2)
+\frac 1 {2z}\zb (0) +\frac 1 z \zb (-1/2) +\Gamma(z)\zeta_{\can+1}(z)\right)
.\nn
\eeq
Several nonlocal pieces, difficult to determine, have cancelled between 
$Z'(0)$ and the $A_i (s)$. 

The small $z$ expansion, 
\beq
\Gamma(z)\zeta_{\can+1}(z) \sim\frac 1 z \zbe (0) -\gamma \zbe (0) 
+\zbe ' (0),\nn
\eeq
must now be employed where
the value of $\zbe (0)$ follows from the fact, \cite{voros87}, that
it equals the $C_D$ term in the asymptotic $t\to 0$ expansion of
\beq
\sum d(\nu ) \frac{e^{-t\nu}}{e^t-1} =\sum_{n=0} ^{\infty} C_n t^{n-D},\nn
\eeq
and can be found using (\ref{9.3}). Explicitly
\beq
\zbe (0) =-\frac 1 2 \zb (0) -\zb (-1/2) 
-2\sum_{i=1}^d \rzb (i/2) \frac{\zeta_R (-i)} i.\label{null}
\eeq
Using the above results and notation, the derivative emerges in the
final form
\beq
\zeta_{\cam} ' (0) &=& \zbe ' (0) + \ln2\bigg(\zb (-1/2)
+2\sum_{i=1}^d\rzb (i/2)D_i (1)\bigg) 
\label{9.6}\\
& &+2\sum_{i=1}^d \rzb (i/2) \left(D_i(1)\sum_{k=1}^{i-1}
\frac 1 k 
+\int_0^1{D_i(t)-tD_i(1)\over t( 1-t^2)}\,dt
\right)\nn
\eeq
after some minor manipulation.
It is seen that all $\gamma$-dependent pieces have cancelled and, 
in short, apart from contributions coming from the $\zbe (z)$ piece,
the functional determinant is determined through the leading
heat-kernel coefficients on the manifold $\can$.

It does not seem possible to proceed much further for the general case because
there is no explicit expression for $\zbe (z)$.
The best we have found is the integral representation
\beq
\zeta_{\can+1}(z) =\frac 1 {2\pi i} \int_{c-i\infty}^{c+i\infty} ds\,\,
{\Gamma (s) \Gamma (z-s)\over\Gamma(z)}\,\zeta_R (z-s)\,\zb (s/2),
\eeq
with $\Re c > d$, which one may find starting from (\ref{nzet}) using the
Mellin Barnes integral representation of the exponential function. 
Equation (\ref{null}) is recovered in this representation closing the contour
to the left.

However, for the example of the monopole, one can continue directly, as will 
be shown in the next section. 

\s{Monopole determinant.}
A situation of possible physical significance is the global monopole. In the
infinite case, Mazzitelli and Lousto 
\cite{mazzitellilousto91} have evaluated some local vacuum
averages on $\reals\times\man$. In the bounded case we can find the effective 
one-loop action on $\man$ in the guise of the functional determinant and so 
we now specialise $\man$ to be the global monopole of section 5. 

For conformal coupling we had
\beq
\nu = \frac 1 a \left( l+\frac{d-1} 2 \right)\nn
\eeq
and the base zeta function, eq.(\ref{4.7}), is
\beq
\zb (s) = a^{2s} \sum_{l=0}^{\infty} d(l) \left( l+\frac{d-1} 2 \right)^{-2s}
.\nn
\eeq
Furthermore we have
\beq
\zbe (s)& =& a^s \sum_{n=1}^{\infty} \sum_{l=0}^{\infty}
 d(l) \left( l+an +\frac{d-1} 2\right)^{-s} \label{10.1}\\
&=& a^s\bigg(\zeb\big(s,(d+1)/2+a|\vec r\big) +\zeb\big(s,(d-1)/2 
+a|\vec r\big)\bigg),\nn
\eeq
with $\vec r =(\vec 1,a)$, $\vec 1$ being the $d$-dimensional unit vector.
Thus, together with eq.~(\ref{9.6}), the functional determinant has been
found 
in terms of derivatives of Barnes zeta functions and a given polynomial in the
radius $a$ and the dimension $d$. The polynomial follows from 
eqs.~(\ref{4.13}) and (\ref{4.12}),
\beq
\zb (0) &=& -\frac{2^{1-d}}{(d-1)d!}\,\pold _d,\nn\\
\zb (-1/2)&=& \frac{2^{1-d}}{a(d-1)(d+1)!}\,\pold _{d+1}.\nn
\eeq

For arbitrary radius $a$ it seems that one cannot construct the  
analytical continuation needed to find an explicit expression for
$\zbe ' (0)$. A numerical treatment could start immediately from 
eqs.~(\ref{9.1}) and (\ref{9.2}) or from the formulas in section 39 of ref.
\cite{barnes03}. For a rational radius one can go 
further, as explained for example in \cite{dowkerapps95}, however
we proceed here only for the ball, {\it i.e.} $a=1$. Then, the zeta 
function,
eq.~(\ref{10.1}), may be expressed in terms of just Hurwitz zeta functions in
the following way.

First one finds
\beq
\zbe (s) =\sum_{l=0} ^{\infty} e(l) \left(l+\frac{d+1} 2 \right)^{-s}
\nn
\eeq
with the ``degeneracy"
\beq
e(l) =( 2l+d) \frac{(l+d-1)!}{l!\,d!}.\nn
\eeq
Then, expanding $e(l)$ in standard fashion as
\beq
e(l) =\sum_{\al =0}^d e_{\al} \left(l+\frac{d+1} 2 \right) ^{\al},\nn
\eeq
the representation
\beq
\zbe (s) = \sum_{\al =0}^d e_{\al}\, \zeta_H\big(s-\al; (d+1)/2 \big) 
\label{10.2}
\eeq
in terms of the Hurwitz zeta function, $\zeta_H $, follows. So finally 
({\it cf} \cite{barnes203} p432)
\beq
\zbe' (0) = \sum_{\al =0}^d e_{\al}\, \zeta_H'\big(-\al; (d+1)/2\big). 
\label{10.3}
\eeq

All quantities needed to calculate the functional determinant on the ball
are now provided. The results agree with previous ones presented
in \cite{bgek,dowkerapps95,dowodd,dowker96} for dimensions $D$ from 2 to 8. 
The structure of those results, such as the absence of
the constant $\gamma$, the appearance of the derivatives of Riemann zeta 
functions with arguments up to $-d $ and a certain prefactor of the 
$\ln 2$ term, is made completely clear with eqs.~({\ref{9.6}) and 
(\ref{10.3}) and is now shown to be true for all dimensions $D$.
\s{Robin functional determinants.}
Let us describe briefly the analogous treatment for Robin boundary 
conditions. Having the comments at the end of section 3 in mind, the 
contributions coming from the $A_i $'s are given by eq.~(\ref{9.1}) with 
the mentioned replacements. Following once more the lines of \cite{bgek}
we find 
\beq
Z_R '(0) = Z'(0)
+N(u),\label{11.1}
\eeq
with $N(u)$ given by
\beq
N(u)
=\sum d(\nu ) \left(
-\ln\left( 1+\frac u {\nu} \right)
+\sum_{n=1}^d (-1)^{n+1} \frac 1 n \left( \frac
u {\nu}\right)^n\right),\nn
\eeq
and $u=1-D/2-\ga$. Thus for Robin conditions we have to treat
only one additional piece, the last one in eq.~(\ref{11.1}), in order to reach
the result analogous to eq.~(\ref{9.6}) for Dirichlet conditions.

To proceed, write $N(u)$ in the form
\beq
N(u)=\sum d(\nu ) \int_0^{\infty}dt\, \frac{e^{-\nu t}} t 
\left(e^{-ut}+\sum_{n=0}^d (-1)^{n+1} \frac{u^nt^n}{n!}\right),\nn
\eeq
and again introduce a regularization parameter 
$z$, as in the derivation of eq.~(\ref{9.4}). 
We find for the resulting function, $N(u,z)$,
\beq
N(u,z) =\ze_\can(z,u)\,\Gamma (z) +\sum_{n=0}^d
(-1)^{n+1} \frac{u^n}{n!}\Gamma (z+n)\,\zb((z+n)/2),\nn
\eeq
where we have introduced
\beq
\ze_\can (z,u) =\frac 1 {\Gamma (s)}
\sum d(\nu ) \int _0^{\infty} dt\,\, t^{z-1} e^{-(\nu +u)t}.\nn
\eeq
One easily obtains as explained previously
\beq
\ze_\can(0,u) = \zb (0) +2\sum_{l=1}^d (-1)^{l} \frac{u^{l}} l 
\rzb (l/2),\nn
\eeq
which guarantees that the limit $z\to 0$ is well defined. These relations 
lead to
\beq
N(u) &=& 
\ze'_\can  (0,u) -\frac 1 2 \zb ' (0) \nn\\
& &+\sum_{n=1}^d (-1)^{n+1} \frac{u^n} n 
\bigg(2\rzb (n/2) \left(\psi (n) +\gamma\right) +\pzb (n/2)\bigg)
.\nn 
\eeq

As in  Dirichlet conditions, on adding up all
contributions to the required derivative, several pieces cancel leaving the 
final compact form
\beq
{\zeta_{\cam}^R}' (0) &=& \zbe ' (0) +  \ze'_\can (0,u) 
+\ln2\bigg(\zb (-1/2) 
+2 \sum_{i=1\atop i\,\,odd}^d \rzb (i/2) M_i (1)\bigg)
\nn\\
& &+2\sum_{i=1\atop i\,\, odd}^d \rzb (i/2)  
\bigg(M_i(1)\sum_{k=1}^{i-1} \frac 1 k +
\int_0^1{M_i(t)-tM_i(1)\over t(1-t^2)}dt\bigg)
\label{11.2}\\
& &
+2\sum_{i=1\atop i\,\, even}^d \rzb (i/2)  
\bigg(M_i(1)\sum_{k=1}^{i-1} \frac 1 k +
\int_0^1{M_i(t)-t^2M_i(1)\over t(1-t^2)}dt\bigg)
\nn
\eeq
where $M_i(1)=D_i(1)+(-1)^{i+1}u^i/i$.

This completely parallels eq.~(\ref{9.6}) for Dirichlet conditions.
Again, the $\gamma$-dependence has disappeared, and the nonlocal pieces are
clearly confined to the first two terms which have to be seen
as special functions as they stand. Nothing more can be said without
specializing to simple manifolds.

Let us briefly describe the simplifications occurring for the monopole. All 
pieces are known from the Dirichlet case apart from
\beq
\ze_\can (s,u)& =&a^{s} \bigg(
\zeb \left( s,(d+1)/ 2 +au\right) +\zeb \left( s,(d-1)/2 +au\right)
\bigg)\nn\\
&=& a^{s} \sum_{l=0}^{\infty} d(l) \left(l+(d-1)/2 +au\right)^{-s}.
\nn
\eeq
However, using the procedure explained at the end of section 10, we expand
\beq
d(l) =\sum_{\al =0}^{d-1} e_{\al} (au) \left(l+(d-1)/2 +au\right)
^{\al},\nn
\eeq
and then, since the $e_{\al} (au)$ are polynomials in $au$,
$\ze_\can(s,u)$ appears as a sum of Hurwitz zeta functions 
\beq
\ze_\can(s,u) =a^{s} \sum_{\al =0}^{d-1} e_{\al} (au)\, \zeta_H \left(
s-\al ; (d-1)/2 +au\right) ,\nn
\eeq
its derivative at $s=0$ being
\beq
\ze'_\can(0,u)= \sum_{\al =0}^{d-1} \left(\zeta_H ' (-\al; 
(d-1)/2 +au)-\ln a\,\frac{ B_{\al +1} ((d-1)/2 +au )}{\al +1}\right),
\eeq
where the $B_n(x)$ are ordinary Bernoulli polynomials.

Thus, in this case also, the only contribution not readily available
for arbitrary radius $a$ is the $\zbe$ one. As in Dirichlet conditions, the 
ball case, $a=1$, is easily extracted. Again, the structure of the result 
is completely clear from equation (\ref{11.2}) and the special cases of 
references \cite{bgek,dowker96,dowodd} are very quickly reproduced.

\s{Conclusion.}
Our basic results are embodied in eqs.~(\ref{4.4}), (\ref{9.6}) and 
(\ref{11.2}). The general form of the determinants agrees with the more
special expressions announced in \cite{dowkerl}.

The techniques described here have certain technical and aesthetic 
advantages. For example,
Levitin determined the heat-kernel coefficients on the $D$-ball in terms
of polynomials in $d$ by fitting their general forms 
using values calculated for specific dimensions. Choosing $\can$ to be the 
unit $(D-1)$-sphere in the preceding formalism has allowed us to find these 
polynomial expressions directly and much more quickly. It takes two minutes
using Mathematica to evaluate the first ten polynomials. A similarly rapid
computation holds for the determinants.

The method is clearly capable of being 
applied to other situations. One may wish to change the base $\can$ or to 
choose a different field for physical or for mathematical reasons.

A generalisation of a slightly different character would be to replace the 
metric (\ref{2.1}) by 
$ds^2=dr^2+f(r^2)d\Si^2$ when one would be obliged to analyse the asymptotic
behaviour of the new radial eigensolutions. A particularly
important example is the spherical suspension, $ds^2=d\th^2+\sin^2\th d\Si^2$,
$0\le\th\le\th_0$.
The asymptotic properties of the resulting Legendre functions derived by 
Thorne \cite{thorne57}
have already been  employed by Barvinsky {\it et al} 
\cite{barvinskykamenshchikkarmazin92} in a 
calculation of a one-loop effective action in quantum cosmology.

We reserve for later, expositions of some of these extensions.
\section*{Acknowledgments}
This investigation has been supported by the DFG under the contract number
Bo 1112/4-1.

\appendix
\app{Heat-kernel polynomials for the monopole with
Dirichlet boundary conditions} 
In this appendix we list the leading 
heat-kernel coefficients for the monopole with
Dirichlet boundary conditions. From (\ref{4.15}),
\beq
\facb\hem _{1/2} &=& -{1\over 4}
\nn\\ 
\fac\hem _1 & =& {{5\,d-3}\over {12}} - {d - 3\over {12\,{a^2}}}
\nn\\ 
\facb\hem _{3/2} & = & {(2-d)(5d-4)\over128} + 
   {(d-1)(d-3)\over {48\,{a^2}}}
\nn\\ 
\fac\hem _2 & = & {{-7875 + 14447\,d - 7293\,{d^2} + 1105\,{d^3}}\over {30240}} + 
   {(d-1)(d-5)(5d-3)\over {1440\,{a^4}}} 
\nn\\ & &- 
   {(d-1)(d-3)(5d-13)\over {144\,{a^2}}}
\nn\\ 
\facb\hem _{5/2}& =& {-4992 + 9552\,d - 5692\,{d^2} + 1356\,{d^3} - 
       113\,{d^4}\over {49152}}
\nn\\
& & - 
   {(d-1)(d-3)(d-5)(5d-3)\over {5760\,{a^4}}} 
\nn\\
& &+ 
   {(d-1)(d-3)(d-4)(5d-14)\over {1536\,{a^2}}}
 \nn\\
\fac\hem _3 & = & {{-28999971 + 57597489\,d - 38150066\,{d^2}}\over
{51891840}} 
\nn\\
& &+ {{ 
       11356742\,{d^3} - 1573675\,{d^4} + 82825\,{d^5}}\over {51891840}} 
\nn\\ & &- 
   {(d-1)(d-3)(d-7)(35d^2-28d+9\over
       {362880\,{a^6}}} 
\nn\\
& &+ {(d-1)(d-3)(d-5)(5d-3)(5d-23)\over {17280\,{a^4}}} 
\nn\\
& &- 
   {(d-1)(d-3)(1105d^3-13923d^2+568789d-74781\over {362880\,{a^2}}}.
\nn
\eeq                
\app{Heat-kernel polynomials for the monopole with
Robin boundary conditions} 
The following is the list for Robin boundary conditions:
\beq
\facb\hem _{1/2}& =& {1\over 4}
\nn\\ 
\fac\hem_1 & =& {{-3 + 5\,d + 24\,\ga}\over {12}} + 
   {{3 - d}\over {12\,{a^2}}}
\nn\\ 
\facb\hem_{3/2}& =& {{8 - 10\,d + 7\,{d^2} + 32\,d\,\ga + 
       64\,{{\ga}^2}}\over {128}} - 
   {(d-1)(d-3)\over {48\,{a^2}}}
\nn\\ 
\fac\hem_2 & = & {{1035 - 871\,d - 75\,{d^2} + 295\,{d^3} + 2160\,\ga - 
       2304\,d\,\ga }\over {4320}} 
\nn\\
& &+ {{ 
2448\,{d^2}\,\ga + 
       5760\,d\,{{\ga}^2} + 5760\,{{\ga}^3}}\over {4320}} 
\nn \\
& &+ 
   {(d-1)(d-5)(5d-3)\over {1440\,{a^4}}} 
\nn\\
& &- 
   {(d-1)(d-3) \,
       \left( 11 + 5\,d + 24\,\ga \right) \over {144\,{a^2}}}
\nn\\ 
\facb\hem_{5/2} & = & {{24960 - 31344\,d + 11668\,{d^2} - 2836\,{d^3} + 
       1587\,{d^4} + 30720\,\ga} \over{245760}}  
\nn\\
& &+{{
- 19200\,d\,\ga - 
       3520\,{d^2}\,\ga + 14560\,{d^3}\,\ga + 
       30720\,{{\ga}^2} - 25600\,d\,{{\ga}^2}
}\over {245760}}\nn\\
& & +{{ 
       56320\,{d^2}\,{{\ga}^2} + 92160\,d\,{{\ga}^3} + 
       61440\,{{\ga}^4}}\over {245760}} 
\nn\\
& &+ 
   {(d-1)(d-3)(d-5)(5d-3)\over {5760\,{a^4}}} 
\nn\\
& &+ 
   {(d-1)(d-3)\,
       \left( -56 + 6\,d - 7\,{d^2} - 64\,\ga - 32\,d\,\ga - 
         64\,{{\ga}^2} \right) \over {1536\,{a^2}}}
\nn 
\eeq
\beq
\lefteqn{
\fac\hem_3  = {{1087}\over {1920}} + {{1744109\,{d^5}}\over {259459200}} + 
   {{9\,\ga}\over {16}} + {{{{\ga}^2}}\over 3} + 
   {{{{\ga}^3}}\over 3} + {{8\,{{\ga}^5}}\over {15}}   }
\nn\\
& & + 
   {d^4}\,\,{{\left( -190555 + 4176744\,\ga \right) }\over {51891840}} 
\nn\\
& &+ 
   {d^3}\,\left( -{{1423133}\over {25945920}} - {{2\,\ga}\over {21}} + 
      {{349\,{{\ga}^2}}\over {945}} \right) 
\nn\\
& & + 
   {d^2}\,\left( {{1300721}\over {3706560}} 
      {{293\,\ga}\over {1080}} - {{7\,{{\ga}^2}}\over {45}} + 
      {{31\,{{\ga}^3}}\over {35}} \right)  
\nn\\
& &+ 
   d\,\left( -{{23787571}\over {28828800}} - {{194\,\ga}\over {315}} + 
      {{11\,{{\ga}^2}}\over {945}} - {{8\,{{\ga}^3}}\over {35}} + 
      {{16\,{{\ga}^4}}\over {15}} \right)  
\nn\\
& &- 
   {(d-1)(d-3)(d-7)(35d^2-28d+9)\over
     {362880\,{a^6}}} 
\nn\\
& &+ {(d-1)(d-3)(d-5)(5d-3)\,
       \left( 25 + 5\,d + 24\,\ga \right) \over {17280\,{a^4}}} 
\nn\\
& &+ 
   {(d-1)(d-3)\over{51840\,{a^2}}}\,     
  \left( -10917 + 3367\,d - 603\,{d^2} - 295\,{d^3} 
- 10800\,\ga + 576\,d\,\ga 
\right.\nn\\
& &\vspace{3cm}\left.
- 2448\,{d^2}\,\ga - 
         5760\,{{\ga}^2} - 5760\,d\,{{\ga}^2} - 
         5760\,{{\ga}^3} \right). 
\nn
\eeq

\end{document}